\RequirePackage{fix-cm}
\documentclass[smallextended]{svjour3}
\smartqed
\usepackage[top=2.5cm,right=3cm,bottom=2.5cm,left=2.5cm]{geometry}
\usepackage{graphicx}
\usepackage{caption}
\usepackage{subcaption}
\usepackage{array,multirow}
\usepackage{amssymb, amsmath}
\usepackage{dutchcal}
\usepackage{braket}
\usepackage{calrsfs}
\DeclareMathAlphabet{\pazocal}{OMS}{zplm}{m}{n}%%%%
\usepackage{xfrac}
\newcommand{\nnlosim}{NNLO\textsubscript{sim}}
\newcommand{\nm}{\ensuremath{N_{\mathrm{max}}}}
\newcommand{\Leff}{\ensuremath{L_{\mathrm{eff}}}}
\newcommand{\w}{\ensuremath{\hbar\omega}}
\newcommand{\YN}{\ensuremath{Y\!N}}
\newcommand{\NN}{\ensuremath{N\!N}}
\newcommand{\piN}{\ensuremath{\pi N}}
\newcommand{\NNN}{\ensuremath{N\!N\!N}}
\newcommand{\chieft}{\(\chi\)EFT}
\def\nuc#1#2{\relax\ifmmode{}^{#1}{\protect\mathrm{#2}}\else${}^{#1}\mathrm{#2}$\fi}
\def\lnuc#1#2{\relax\ifmmode{}_{\Lambda}^{#1}{\protect\mathrm{#2}}\else${}_\Lambda^{#1}\mathrm{#2}$\fi}
\newcommand{\LamUV}{\ensuremath{\Lambda_{\mathrm{UV}}}}
\newcommand{\LamNN}{\ensuremath{\Lambda_{\NN}}}
\newcommand{\LamYN}{\ensuremath{\Lambda_{\YN}}}
\newcommand{\Tmax}{\ensuremath{T_{\rm Lab}^{\rm max}}}

\begin{document}

\title{Systematic Nuclear Uncertainties in the Hypertriton System}
\author{Thiri~Yadanar~Htun \and
                Daniel~Gazda  \and
                Christian~Forssén \and
                Yupeng~Yan}

\institute{T.Y.~Htun$^{1,2,4}$ \and D.~Gazda$^{1,3}$ \and C.~Forssén$^1$ \and Y.~Yan$^2$
  \at $^1$Department of Physics, Chalmers University of Technology, SE-412 96 Göteborg, Sweden\\
  $^2$School of Physics and Center of Excellence in High Energy Physics and Astrophysics, Suranaree University of Technology, Nakhon Ratchasima, 30000, Thailand \\
  $^3$Nuclear Physics Institute of the Czech Academy of Sciences, 25068 \v{R}e\v{z}, Czech Republic\\
  $^4$Department of Physics, University of Mandalay, 05032 Mandalay, Myanmar
}

\date{Received: date / Accepted: date}

\maketitle

\begin{abstract}
  The hypertriton bound state is relevant for
  inference of knowledge about the hyperon--nucleon (\YN) interaction.
  In this work we compute the binding energy of the hypertriton using the
  \emph{ab initio} hypernuclear no-core shell model (NCSM) with
  realistic interactions derived from chiral effective field theory.
  In particular, we employ a large family of nucleon--nucleon interactions
  with the aim to quantify the theoretical precision of predicted
  hypernuclear observables arising from nuclear-physics uncertainties.
  The three-body calculations are performed in a relative
  Jacobi-coordinate harmonic oscillator basis and we implement
  infrared correction formulas to extrapolate the NCSM results to
  infinite model space.
  We find that the spread of the predicted hypertriton binding
  energy, attributed to the nuclear-interaction model uncertainty, is
  about 100~keV.
  In conclusion, the sensitivity of the hypertriton binding energy to
  nuclear-physics uncertainties is of the same order of magnitude as
  experimental uncertainties such that this bound-state observable can
  be used in the calibration procedure to constrain the \YN{}
  interactions.
\end{abstract}

\section{Introduction}
\label{intro}
Hypernuclei play an essential role in understanding the interactions
between nucleons and hyperons. Unlike the case of nuclear
interactions---with a vast database of precisely measured low-energy
nucleon--nucleon (\NN{}) observables---the available experimental data
on (\YN) scattering are much poorer both in quantity and quality.

Fortunately, there is an immense amount of information on \YN{}
interactions encoded in precision measurements of hypernuclear
properties, such as the \(\Lambda\) hyperon separation energies and
excitation spectra, which has been collected over the past
decades~\cite{Davis:2005mb}. However, the utilisation of precise
experimental data on hypernuclear spectroscopy to constrain the
underlying \YN{} interactions is a very challenging task. First, the
computationally-demanding \emph{ab~initio} methods, which do not rely
on uncontrolled approximations to renormalize the hypernuclear
interactions, started to emerge only
recently~\cite{Nogga:2013pwa,Lonardoni:2013rm,Wirth:2014apa,Wirth:2017bpw,Contessi:2018qnz,Le:2020zdu,Schafer:2021mit}.
Second, it is very important to understand all sources of
uncertainties in the calculations to quantify the theoretical
precision of relevant observables. The main source of uncertainty in
\emph{ab~initio} calculations of light hypernuclei is the hypernuclear
Hamiltonian, constructed from particular models of nuclear and
hypernuclear interactions. Besides the poorly constrained \YN{}
interactions, the remaining freedom in the construction of realistic
nuclear forces represents an additional source of model uncertainty
which propagates into hypernuclear observables. In recent years, there
has been significant progress in the uncertainty quantification in
nuclear
forces~\cite{Furnstahl:2015rha,Ekstr_m_2015,Carlsson:2015vda,P_rez_2015,PhysRevC.93.044002},
which allowed to provide theoretical uncertainties in nuclear
structure
calculations~\cite{Carlsson:2015vda,PhysRevC.93.044002,Perez:2015bqa,Acharya:2016kfl}.

In this work, we study the sensitivity of the hypertriton binding
energy to systematic model uncertainties in the nuclear interactions.
This information is very important for the future use of this
observable to constrain the \YN{} interactions. More specifically we
perform hypernuclear NCSM calculations in relative Jacobi-coordinate
harmonic oscillator (HO) basis with a family of 42 realistic chiral
\NN{} interactions (introduced in Sec.~\ref{sec:unc}) while keeping
the \YN{} interaction fixed. We study in detail the convergence
properties of the NCSM calculations and the dependence of observables
on the NCSM model-space parameters. Moreover, we apply infrared (IR)
correction formulas~\cite{Wendt:2015nba} to extrapolate the results
obtained in finite model spaces to infinite model space, in the
hypernuclear NCSM. We present results for the hypertriton ground-state
energy and discuss the consequences of our findings. In general, this
study serves as a starting point for the uncertainty quantification of
other hypernuclear observables.

\section{Methodology}
\subsection{Three-body system in the hypernuclear no-core shell model}
\label{sec:ncsm}
The starting point to study the hypertriton within the hypernuclear
NCSM approach~\cite{Wirth:2017bpw} is the nonrelativistic Hamiltonian
for three particles interacting by realistic \YN{} and \NN{}
interactions
\begin{equation}
  \label{eq:h}
  H = -\sum_{i=1}^3\frac{\hbar^2}{2m_i}\vec{\nabla}^2_i +
  V_{\NN}(\vec{r}_1,\vec{r}_2) + \sum_{i=1}^2 V_{\YN{}}(\vec{r}_i,
  \vec{r}_3) + \Delta M,
\end{equation}
where the coordinates \(\vec{r}_i\) and masses \(m_i\) correspond to
nucleons for \(i=1,2\) and for \(i=3\) to a hyperon. Since we
explicitly take into account the strong-interaction
\(\Lambda N \leftrightarrow \Sigma N\) transitions in
\(V_{\YN}\)~\cite{Polinder:2006zh}, the \(\Lambda\)-hypernuclear
states are coupled with \(\Sigma\)-hypernuclear states. To account for
the mass difference of these states, the mass term \(\Delta M\) is
introduced in Eq.~\eqref{eq:h}.

The formulation of NCSM which is particularly suitable for few-body
hypernuclear systems employs a translationally-invariant HO basis
defined in relative Jacobi coordinates and the associated momenta. For
an \(A=3\) hypernucleus we need to introduce two different sets of
Jacobi coordinates. The first set, expressed in term of rescaled
single-particle coordinates \(\vec{x}_i=\sqrt{m_i}\,\vec{r}_i\), is
defined as
\begin{equation}
  \label{eq:xi}
  \begin{split}
    \vec{\xi}_0 &= \frac{1}{\sum_{i=1}^3m_i} \sum_{i=1}^3 \sqrt{m_i} \vec{x}_i,\\
    \vec{\xi}_{1} &=\sqrt{\frac{1}{2}}\left(\vec{x}_1 - \vec{x}_2\right),\\
    \vec{\xi}_{2} &=\sqrt{\frac{2m_N m_Y}{2m_N+m_Y}}\left[\frac{1}{2\sqrt{m_N}}(\vec{x}_1+\vec{x}_2)-\frac{1}{\sqrt{m_Y}}\vec{x}_3\right],\\
  \end{split}
\end{equation}
where \(m_N\) and \(m_Y\) are the nucleon and hyperon (\(\Lambda\) or
\(\Sigma\)) masses. In this set, the coordinate \(\vec{\xi}_0\) is
proportional to the center of mass (c.m.) coordinate of the \(A=3\)
system, \(\vec{\xi}_1\) is proportional to the relative coordinate of
the nucleons, and \(\vec{\xi}_2\) is proportional to the relative
coordinate of the hyperon with respect to the c.m.\ coordinate of the
nucleon pair. Since the interaction terms in the
Hamiltonian~\eqref{eq:h} depend only on the relative coordinates, the
center of mass motion, associated with \(\vec{\xi}_0\), can be
eliminated to decrease the dimensionality of the problem.
Consequently, the intrinsic wave function of the \(A=3\) system can be
expanded in a complete set of HO basis states
\begin{equation}
  \label{eq:basis1}
  \ket{(n_{\NN}(l_{\NN} s_{\NN})j_{\NN}t_{\NN}, \pazocal{N}_Y \mathcal{L}_Y  \pazocal{J}_Y \pazocal{T}_Y)JT}
\end{equation}
characterized by a single HO frequency \w, where
\(\ket{n_{NN}(l_{NN} s_{NN})j_{NN}t_{NN}}\) and
\(\ket{\pazocal{N}_Y \mathcal{L}_Y \pazocal{J}_Y \pazocal{T}_Y}\) are
HO states, depending on the coordinates \(\vec{\xi}_1\) and
\(\vec{\xi}_2\), respectively. The \(n_{\NN} (\pazocal{N}_Y)\),
\(l_{\NN}(\pazocal{L}_Y)\), \(s_{\NN}(\tfrac{1}{2})\),
\(j_{\NN}( \pazocal{J}_Y)\), \(t_{\NN}(\pazocal{T}_Y)\) are the
radial, orbital, spin, angular momentum, and isospin quantum numbers
corresponding to the relative two-nucleon (hyperon) state. The \NN{}
and hyperon (\(Y=\Lambda,\Sigma\)) states are coupled to the total
angular momentum \(J\) and isospin \(T\). Antisymmetry of the
states~\eqref{eq:basis1} with respect to nucleon interchange is
achieved by restricting the two-nucleon channel quantum numbers by the
condition \((-1)^{l_{\NN}+s_{\NN}+t_{\NN}}=-1\). The set of Jacobi
coordinates~\eqref{eq:xi} is convenient for the construction of
antisymmetric translationally-invariant HO basis and evaluation of the
\NN{} interaction matrix elements. It is not, however, suitable for
the evaluation of the \YN{} interaction matrix elements.

In order to evaluate the \(V_{\YN}\) matrix elements, another set of
Jacobi coordinates is needed. The new set is obtained from the
set~\eqref{eq:xi} by keeping the coordinate \(\vec{\xi}_0\) and
introducing two new coordinates
\begin{equation}
  \label{eq:eta}
  \begin{split}
    \vec{\eta}_{1} &=\sqrt{\frac{(m_N+m_Y)m_N}{2m_N+m_Y}}\left
      [\frac{1}{\sqrt{m_N}}\vec{x}_1-
      \frac{1}{(m_N+m_Y)}(\sqrt{m_N}\vec{x}_2+\sqrt{m_Y}\vec{x}_3)\right],\\
    \vec{\eta}_{2} &=\sqrt{\frac{m_N m_Y}{m_N+ m_Y}}\left(\frac{1}{\sqrt{m_N}}\vec{x}_2 -\frac{1}{\sqrt{m_Y}}\vec{x}_3\right),
  \end{split}
\end{equation}
where \(\vec{\eta}_1\) is the relative coordinate of a nucleon with
respect to the c.m.\ of the \YN{} pair and \(\vec{\eta}_2\) is the
relative coordinate of the \YN{} pair. The Jacobi coordinates in this
set can be obtained by an orthonormal transformation of the
coordinates~\eqref{eq:xi}. Consequently, the HO
states~\eqref{eq:basis1} can be expanded as
\begin{equation}
  \label{eq:expansion}
  \begin{split}
    &\ket{(n_{\NN}(l_{\NN}s_{\NN})j_{\NN}t_{\NN} ,\pazocal{N}_Y \mathcal{L}_Y \pazocal{J}_Y \pazocal{T}_Y)JT}
    =\sum_{LS}\widehat{L}^2\widehat{S}^2
    \widehat{j}_{NY}\widehat{\pazocal{J}}_{N}\widehat{j}_{\NN}\widehat{\pazocal{J}}_Y
    \widehat{s}_{\NN} \widehat{s}_{NY} \\
    &\times (-1)^{s_{NY}+ \sfrac{1}{2}+s_{\NN}+ \sfrac{1}{2} }
    \begin{Bmatrix}
      l_{NY} & s_{NY} & j_{NY}\\
      \mathcal{L}_{N} & \sfrac{1}{2} & \pazocal{J}_{N}\\
      L & S & J
    \end{Bmatrix}
    \begin{Bmatrix}
      l_{\NN} & s_{\NN} & j_{\NN}\\
      \mathcal{L}_Y  & \sfrac{1}{2} & \pazocal{J}_Y\\
      L & S & J
    \end{Bmatrix}
    \begin{Bmatrix}
      \sfrac{1}{2} & \sfrac{1}{2} & s_{\NN}\\
      \sfrac{1}{2} & S & s_{NY}
    \end{Bmatrix}   \\
    &\times(-1)^{t_{NY}+\pazocal{T}_N+t_{NN}+\pazocal{T}_Y} \widehat{t}_{NY}\widehat{t}_{\NN}
    \begin{Bmatrix}
      \sfrac{1}{2} & \sfrac{1}{2} & t_{\NN}\\
      t_Y &  T  & t_{NY}
    \end{Bmatrix} (-1)^{\mathcal{L}_{N}+\mathcal{L}_Y}\\
    &\times
    \braket{n_{NY}l_{NY}\pazocal{N}_N\mathcal{L}_N \vert n_{\NN}l_{\NN}\pazocal{N}_Y \mathcal{L}_Y}_{\frac{2m_N+m_Y}{m_Y}}
    \ket{(n_{NY} (l_{NY}s_{NY})j_{NY}t_{NY},\pazocal{N}_N\mathcal{L}_N\pazocal{J}_N)JT}
  \end{split}
\end{equation}
in terms of HO basis states
\begin{equation}
\ket{(n_{NY} (l_{NY}s_{NY})j_{NY}t_{NY},\pazocal{N}_N\mathcal{L}_N\pazocal{J}_N)JT},
\label{eq:basis2}
\end{equation}
where \(\ket{n_{NY} (l_{NY}s_{NY})j_{NY}t_{NY}}\) and
\(\ket{\pazocal{N}_N\mathcal{L}_N\pazocal{J}_N}\) are HO states
corresponding to the nucleon--hyperon pair and a nucleon and depending
on the coordinates \(\vec{\eta}_2\) and \(\vec{\eta}_1\),
respectively. In Eq.~\eqref{eq:expansion},
\(\widehat{j} = \sqrt{2j+1}\), the terms in curly brackets are Wigner
6-{} and 9-\(j\) symbols, and \(\braket{\cdot \vert \cdot}_d\) is the
general HO bracket for two particles with mass ratio
\(d=\frac{2m_N+m_Y}{m_Y}\)~\cite{Kamuntavicius:2001pf}, which
facilitates the transformation between coordinates
\(\vec{\xi}_1, \vec{\xi}_2\) and \(\vec{\eta}_2, \vec{\eta}_1\). With
the help of expansion~\eqref{eq:expansion}, it is straightforward to
evaluate the \(V_{\YN}\) matrix elements as
\begin{equation}
  \label{eq:vny}
  \braket{\sum_{i=1}^2 V_{\YN{}}(\vec{r}_i,
  \vec{r}_3)} = 2 \braket{V_{\YN}(\vec{\eta}_2)},
\end{equation}
where the matrix element on the right hand side is diagonal in all
quantum numbers of the states~\eqref{eq:basis2}, except for \(n_{NY}\)
and \(l_{NY}\).

NCSM calculations are performed in finite model spaces by
diagonalization of the matrix representation of the
Hamiltonian~\eqref{eq:h}. The model space is truncated by restricting
the maximum number of HO quanta in the basis states~\eqref{eq:basis1}
as
\begin{equation}
  \label{eq:trunc} 2\,n_{NN}+l_{NN}+2\,\pazocal{N}_Y+\mathcal{L}_Y
  \leq \nm.
\end{equation}
The NCSM calculations are thus variational and converge to exact
results for \(\nm \to \infty\).

In this work, we employed the leading-order (LO) Bonn--J\"{u}lich
SU(3)-based \chieft{} \YN{} model~\cite{Polinder:2006zh} with
regulator cutoff momentum \(\LamYN=600\)~MeV, together with the
\nnlosim{} family of chiral nuclear interactions at
next-to-next-to-leading-order (NNLO) (for details see
Sec.~\ref{sec:unc}). No renormalization was applied to either of the
interactions.

\subsection{Infrared Length Scale of the NCSM basis and Extrapolations}
IR extrapolation can be used to estimate the infinite-space limit from
results computed in truncated model
spaces~\cite{Furnstahl:2012qg,Coon:2012ab}. The truncation of the HO
basis in terms of model space (\nm) and frequency (\w) can be
translated into associated IR and ultraviolet (UV) scales. In
particular, for the NCSM basis that is associated with a total energy
truncation the corresponding IR scale, \Leff{}, can be extracted by
studying the discrete kinetic energy spectrum~\cite{Wendt:2015nba}.
The LO IR extrapolation formula for energies
is~\cite{Furnstahl:2012qg}
\begin{equation}
  E(\Leff)=E_\infty+a_0 e^{(-2k_\infty \Leff)},
  \label{eq:E_LO_IR}
\end{equation}
where $E_\infty$, $a_0$ and $k_\infty$ are the fit parameters.
Subleading IR corrections have the expected
magnitude~\cite{Forssen:2017wei}
\begin{equation}
  \sigma_\text{IR} \propto \frac{e^{-2k_\infty \Leff}}{k_\infty \Leff}.
  \label{eq:E_NLO_IR}
\end{equation}

In addition, UV corrections to finite-space results can be significant
unless $\LamUV \gg \LamNN,\LamYN$. The IR extrapolation
formula~\eqref{eq:E_LO_IR} is still valid at a fixed UV
scale~\cite{Forssen:2017wei}. Such data can be obtained by performing
computations at appropriate $(\nm,\w)$ model-space parameters. The
extrapolated result will then depend on the selected UV scale. This UV
dependence can be monitored such that a sufficiently large scale is
used to achieve UV-convergence. In this work we find that
$\LamUV=1200$~MeV is sufficient. A more detailed study of IR-,
UV-scales and the performance of extrapolation methods in hypernuclear
many-body systems will be the topic of future work. In this study we
have specifically used IR extrapolation as an independent check of the
convergence of the variational minimum.
\begin{figure}
  \centering
  \includegraphics[width=0.64\textwidth]{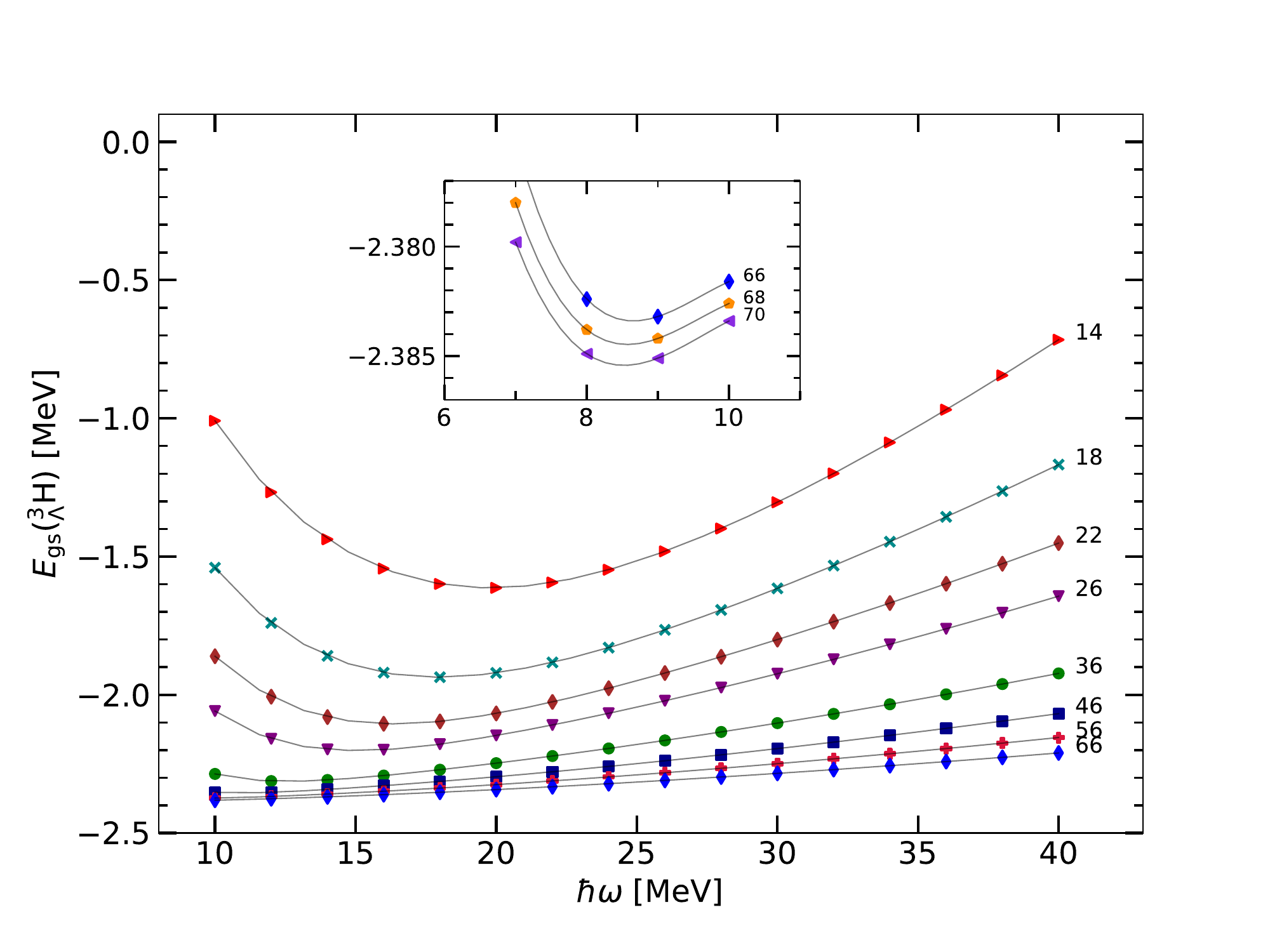}
  \caption{The \lnuc{3}{H} g.s.\ energy \(E(\lnuc{3}{H})\) as a function
    of the HO frequency \w{}, calculated using the \nnlosim{}
    Hamiltonian with $\LamNN=500$~MeV and $\Tmax=290$~MeV for several
    model-space truncations from \(\nm=14\) to \(\nm=70\).%
    \label{fig:Econvergence}%
  }
\end{figure}

\subsection{Systematic Uncertainties of Nuclear Interactions}
\label{sec:unc}
The chiral \NN{} interaction model is associated with systematic
uncertainties that originate in the selection of calibration data, the
truncation of the chiral expansion and possible regulator artefacts.
Here we quantify the magnitude of these systematic uncertainties by
employing the full family of 42 different interactions at NNLO
(labeled \nnlosim) that was constructed by Carlsson et
al.~\cite{Carlsson:2015vda}. These interactions were obtained using
six different truncations of the \NN{} scattering calibration data
($T_\text{lab} \leq \Tmax$ with $\Tmax \in [125, 290]$~MeV) and seven
different regulator cutoffs
($\LamNN{} \in \{ 450, 475, \ldots, 575, 600\}$~MeV). For each
different interaction model, all 26 low-energy constants (LECs) up to
NNLO were simultaneously optimised to \NN{} and \piN{} scattering data
plus bound-state observables in the few-nucleon sector. The
statistical uncertainties in the LECs from the fit have a small effect
on the computed hypernuclear binding energies. Instead, we focus our
attention on the spread of the predicted hypertriton binding energy
obtained with the 42 different interactions. It should be stressed
that all these interaction models give an equally good description of
the fit data.

\section{Results and Discussion}
NCSM calculations are performed for the hypertriton with model space
truncations $\nm \le 66$, and in the range of HO frequencies
$7 \leq \w \leq 40$~MeV with the chiral \nnlosim{} family of \NN{}
interactions and a fixed LO \YN{} interaction. These are basically 42
independent calculations. The hypertriton ground-state energy as a
function of the model space truncation and HO frequency \w{} is
presented in Fig.~\ref{fig:Econvergence} for one of the 42 \nnlosim{}
Hamiltonians ($\LamNN =500$~MeV, $\Tmax = 290$~MeV).
\begin{figure}
  \centering
\includegraphics[width=0.64\textwidth]{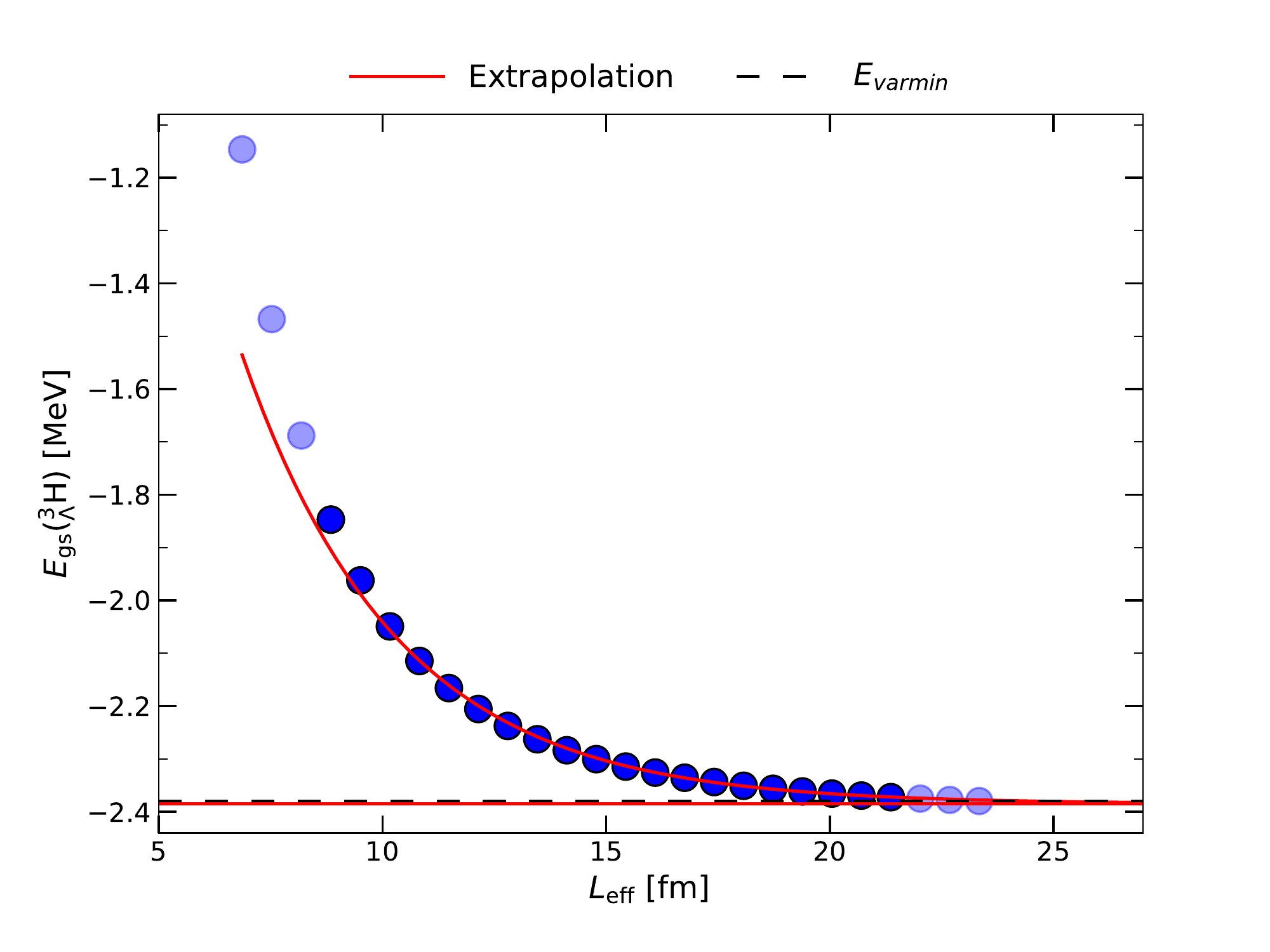}
\caption{%
  Extrapolation of \(E(\lnuc{3}{H})\) calculated using the \nnlosim{}
  Hamiltonian with $\LamNN=500$~MeV and $\Tmax=290$~MeV for a fixed value of
  the HO basis UV cutoff \(\LamUV=1200\)~MeV. The fit is performed
  with data $\nm \in [22,60]$ and compared with the variational
  minimum at \(\nm = 66\). %
  \label{fig:E_IR_extrapolation}%
}
\end{figure}
The energy is clearly converging with increasing model space. The slow
convergence at large frequencies can be attributed to the small
binding energy and long tail of the wave function. Consequently we
also computed the energies at $\w = 7,8,9$~MeV to locate the
variational minimum $E_\mathrm{varmin}$. The lowest energy for this
particular interaction, $E(\lnuc{3}{H}) = -2.385$~MeV, was found at
$\w =9$~MeV and $\nm = 70$.

We further demonstrate the convergence of our NCSM results by applying
IR extrapolation and translating the model space parameters $(\nm,\w)$
into an IR length scale \Leff{} and a UV scale $\LamUV$. Selecting
results obtained with a high UV cutoff $\LamUV=1200$~MeV---such that
computations are UV converged---we perform fits to
Eq.~\eqref{eq:E_LO_IR} with weights proportional to the inverse of the
correction term~\eqref{eq:E_NLO_IR}. The extrapolation shown in
Fig.~\ref{fig:E_IR_extrapolation} is performed with data up to
$\nm=60$, as shown by markers with a black border. The extrapolated
result $E_\infty(\lnuc{3}{H}) = -2.385$~MeV is presented by a red
horizontal line. There is just a few keV difference between
$E_{\mathrm{varmin}}$ at \(\nm = 66\) and $E_\infty$ for this
interaction which indicates a negligible many-body method uncertainty.
The experimental binding energy for the hypertriton is
$E_{\mathrm{{exp.}}}(\lnuc{3}{H}) = -2.35 \pm 0.05$~MeV
\cite{Davis:2005mb}.

\begin{figure}
  \centering
  \includegraphics[width=0.64\textwidth]{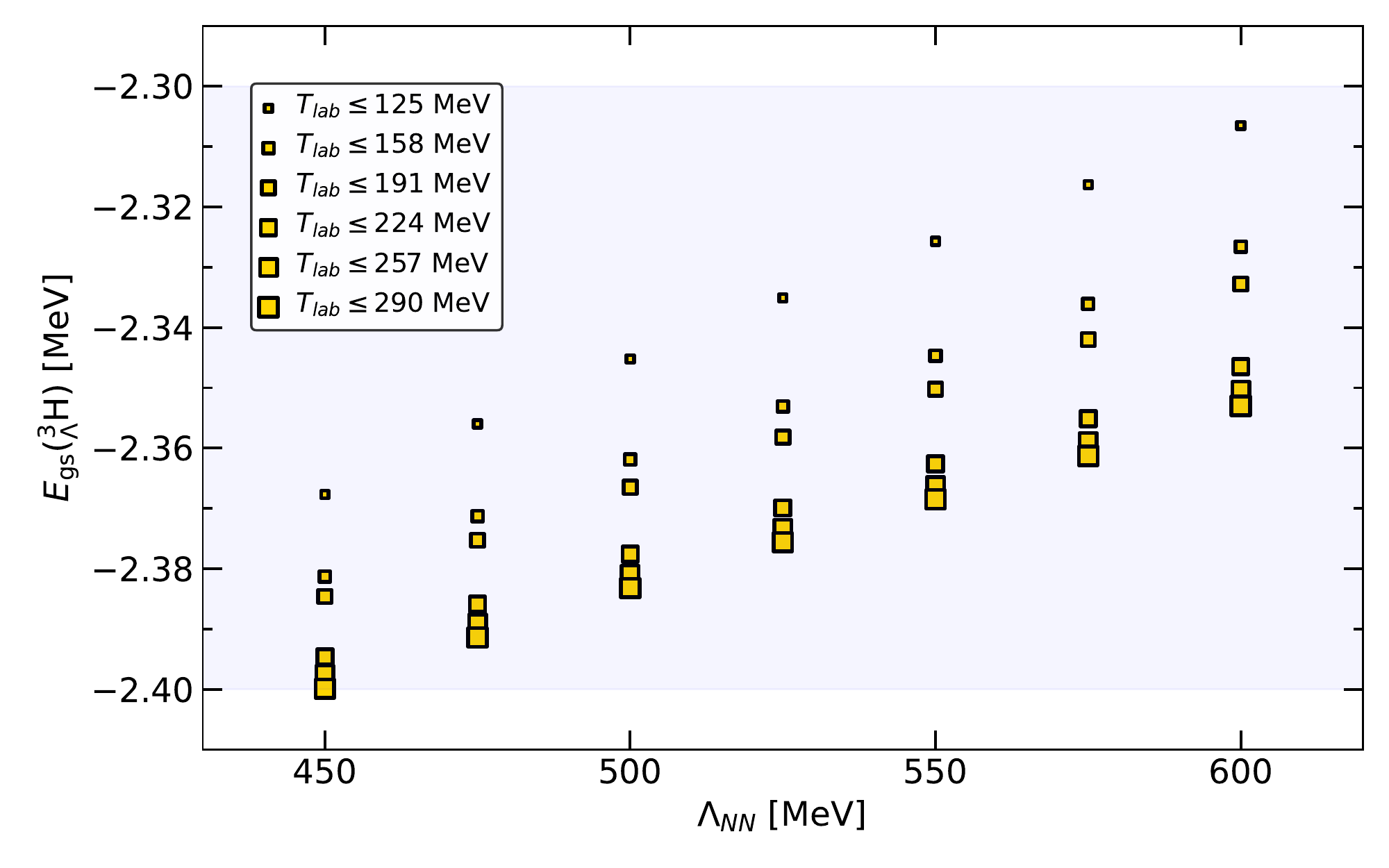}
  \caption{The hypertriton binding energy (variational minima at $\nm
    =66$ and $\w=9$~MeV) for the 42 different \nnlosim{} interaction
    models. The experimental result $E_{\mathrm{{exp.}}}(\lnuc{3}{H}) =
    -2.35 \pm 0.05$~MeV is represented by the blue band.%
    \label{fig:E_NNLOsim}%
  }
\end{figure}

Since we have shown that we can reach converged results (within
$\sim$keV) we perform the computations with all 42 interaction models
using $\nm=66$ and $\w=9$~MeV. The resulting spread in the hypertriton
binding energy is shown in Fig.~\ref{fig:E_NNLOsim}. The predicted
binding energy decreases with increasing regulator cutoff \LamNN{} and
increases with increasing \Tmax. The modest spread in the predictions
indicates that the systematic uncertainty of $E(\lnuc{3}{H})$ due to
the nuclear interaction model is small. The $\lesssim 100$~keV spread
of energies is basically of the same magnitude as the experimental
uncertainty, while the convergence error in the many-body solver is
negligible. The 100~keV spread is small compared to the wide
uncertainty bands that were observed for \nuc{4}{He} and \nuc{16}{O}
with the same family of interactions~\cite{Carlsson:2015vda}. This
difference can largely be attributed to the fact that the \lnuc{3}{H}
system involves only two-body \NN{} interactions while the \NNN{}
forces that are involved in \nuc{4}{He} and \nuc{16}{O} calculations
induce much larger uncertainties. This finding of a modest systematic
uncertainty opens up the opportunity to use the hypertriton binding
energy as a relevant observable toconstrain \YN{} interaction models.

\section{Summary and Outlook}
We have estimated the uncertainty in the predicted hypertriton binding
energy coming from the systematic uncertainty of the \NN{}
interaction. The small-magnitude spread ($\lesssim 100$~keV) found
with the \nnlosim{} family of interactions implies that the
hypertriton binding energy provides an important constraint on \YN{}
interaction models. In the future it will be interesting to quantify
uncertainties of other few- and many-body hypernuclear bound-state
observables to identify additional informative constraints.

\begin{acknowledgements}
  The work of T.Y.~Htun was supported by the Royal Golden Jubilee
  Ph.D.\ Program jointly sponsored by Thailand International
  Development Cooperation Agency, International Science Programme
  (ISP) in Sweden, and Thailand Research Fund under Contract
  No.~PHD/0068/2558. The work of D.~Gazda was supported by the Czech
  Science Foundation GA\v{C}R grant No.\ 19-19640S and by the Knut and
  Alice Wallenberg Foundation (PI: Jan Conrad). The work of C.~Forssén
  was supported by the Swedish Research Council (dnr.~2017-04234).
  Some of the computations were performed on resources provided by the
  Swedish National Infrastructure for Computing (SNIC) at C3SE
  (Chalmers) and NSC (Link\"{o}ping).
\end{acknowledgements}

\bibliographystyle{spphys}
\bibliography{bibliography}
\end{document}